\documentstyle[12pt,psfig]{article}
\def\be{\begin{equation}}
\def\ee{\end{equation}}

\def\lsim{\raise0.3ex\hbox{$<$\kern-0.75em\raise-1.1ex\hbox{$\sim$}}}
\def\gsim{\raise0.3ex\hbox{$>$\kern-0.75em\raise-1.1ex\hbox{$\sim$}}}

\def\PL{{ Phys.\ Lett.\ }}
\def\PR{{ Phys.\ Rev.\ }}

\def\PRL{{ Phys.\ Rev.\ Lett.\ }}

\begin{document}


\vskip 1cm

\centerline{\large{\bf The Evolution of Dynamical Screening}}

\medskip

\centerline{\large{\bf in the Parton Cascade Model}}

\vskip 1cm

\centerline{\bf Helmut Satz$^1$ and Dinesh Kumar Srivastava$^{1,2}$}

\bigskip \bigskip

\centerline{1: Fakult\"at f\"ur Physik, Universit\"at Bielefeld,
D-33501 Bielefeld, Germany~}

\medskip

\centerline{2: Variable Energy Cyclotron Centre, 1/AF Bidhan Nagar,
Calcutta 700 064, India}

\vskip 1.5cm

\begin{abstract}

\par

We determine the time evolution of the colour screening mass in high
energy nuclear collisions, as provided by the parton cascade model.
Using our result, we discuss the onset of deconfinement and the onset of
quarkonium suppression in a gene\-ral, not necessarily equilibrated
environment of strongly interacting constituents.

\end{abstract}

\vskip 1.0cm

Statistical QCD predicts that with increasing temperature, strongly
interacting media will undergo a transition from hadronic matter to
quark-gluon plasma. The aim of high energy nuclear collisions is to
study this transition and the resulting new deconfined phase in the
laboratory. However, statistical QCD deals with large systems in thermal
equilibrium, while nuclear collisions provide small, rapidly evolving
systems. A study of the onset of deconfinement in a microscopic,
non-equilibrium space-time picture would therefore be of great help in
understanding if and to what extent theory and experiment can be
expected to meet. The aim of the parton cascade model \cite{GM,G}
is to describe the evolution of a high energy
nucleus-nucleus collision through perturbative partonic interactions
embedded in a relativistic transport theory.
Hence it seems to be a good starting point
for a `dynamic study of deconfinement'.

\medskip

The initial state of two colliding nucleons can be pictured as colliding
beams of confined partons. The incident partons have a distribution in
intrinsic transverse momentum, leading to some average value $\langle
k_T \rangle$, which in turn defines an average transverse parton size
$r_T= 1/\langle k_T \rangle$. When two nuclei collide at high energy, the
transverse parton density increases considerably, since now we have a
superposition of nucleon-nucleon collisions. In this medium, colour
screening will destroy the association of partons to particular hadrons,
since for a sufficiently high density of colour charges, the colour
screening radius becomes much smaller than the typical hadronic scale.
Hence we expect the onset of deconfinement for some characteristic
density or for an equivalent screening scale. In perturbative QCD, the
method to calculate the colour screening mass $\mu$ (the inverse of
the screening radius $r_D$) is well-known; an extension to the
non-equilibrium medium provided by cascade models has also been
proposed \cite{BMW,sspc}. We shall therefore use this method to obtain the
time evolution of the screening mass $\mu(t)$ in the parton cascade
model.

\medskip

Given $\mu(t)$, we have to specify for what value deconfinement sets
in; the parton cascade model itself does not identify such a point.
It just has the primary collisions between the `confined' partons
of the incident nucleons, followed by successive interactions between
primary as well as produced partons and eventually by the hadronisation
of partons according to some assumed scheme. There are two possible ways
to specify deconfinement in such a picture. One can obtain at any given
time the transverse density profile of the parton distribution, and use
the percolation point of partonic discs in the transverse plane to define
the onset of deconfinement \cite{Armesto}--\cite{Satz-T}. Here we
shall follow a somewhat different path, taking as critical screening mass
$\mu_c$ the value obtained from lattice studies. These provide the
temperature dependence of $\mu$ and in particular also its value at the
deconfinement point $T_c$, where one has $\mu_c \equiv \mu(T_c) \simeq
0.4 - 0.6$ GeV \cite{lattice}. It is not obvious that such an
equilibrium value is really applicable to the non-equilibrium situation
provided by the parton cascade model. A good check would be to compare
in this model the transverse parton density at $\mu_c$ to the
percolation value. Work on this is in progress.

\medskip

Let us briefly recall the most important features of the parton cascade
model.
\begin{itemize}
\vspace*{-0.2cm}
\item{The initial nucleus-nucleus system is treated as two colliding
clouds of partons, whose distributions are fixed by the nucleonic parton
distribution functions determined in deep inelastic lepton-nucleon
scattering, and by the nucleon density distributions in the nuclei.}
\vspace*{-0.2cm}
\item{The parton cascade development starts when the initial parton clouds
interpenetrate, and it follows their space-time development due to
interactions. The model includes multiple elastic and inelastic
interactions described as sequences of $2\, \rightarrow \,2$
scatterings, $1\, \rightarrow \, 2$ emissions, and $2 \, \rightarrow \,1$
fusions. It moreover explicitly accounts for the individual time scale
of each parton-parton collision, the formation time of the parton
radiation, the effective suppression of radiation from virtual partons
due to an enhanced absorption probability by others in regions of dense
phase space occupation, and the effect of soft gluon interference in
low energy gluon emission.}
\vspace*{-0.2cm}
\item{Finally, the hadronization in terms of a parton coalescence to
colour neutral clusters is described as a local statistical process
that depends on the spatial separation and colour of the
nearest-neighbour partons~\cite{eg}. These pre-hadronic clusters then
decay to form hadrons.}
\vspace*{-0.2cm}
\end{itemize}
Since we are interested in the early state of the collision when
the strongly interacting matter is still in partonic form, the
hadronisation part of the model will not play a role for our
considerations.

\medskip

To specify the further basis for our considerations, we add some details of 
the parton-parton scattering considered in the parton cascade model
\cite{GM,G}.
The elementary parton scatterings $a+b\, \rightarrow\,c+d$ and the fusion
processes $a+b\,\rightarrow\, c^{*}$ themselves are devided into two
distinct classes:
\begin{enumerate}
\vspace*{-0.2cm}
\item[i)] 
{\em hard} parton collisions with a sufficiently large momentum transfer
$p_\perp^2$ or invariant mass
$\widehat{s}$ to apply perturbative QCD; and 
\vspace*{-0.2cm}
\item[ii)] 
{\em soft} parton collisions with 
low momentum transfer $p_\perp^2$ or invariant mass $\widehat{s}$,
which will be modelled phenomenologically. 
\vspace*{-0.2cm}
\end{enumerate}
\noindent
This division is necessary to regulate the collision integrals which
appear in the transport equations describing the evolution of the 
partonic system and which are singular
for small momentum transfer $Q^2$ in the Born
approximation. The parton-parton cross-section is rendered finite
by invoking an invariant {\em hard-soft division scale} $p_0^2$ such that
the collisions occuring at a momentum scale $Q^2\geq p_0^2$ 
are treated perturbatively, whereas a soft, non-perturbative 
interaction is assumed for those with $Q^2 < p_0^2$.
The total cross-section for collisions between two
partons is then written as
\begin{equation}
\widehat{\sigma}_{ab}(\widehat{s})=
\sum_{c,(d)}
\left\{ \int_0^{p_0^2} dQ^2 \left( \frac{ d\widehat{\sigma}
^{soft}_{ab\rightarrow c(d)} }{dQ^2}\right)+
 \int_{p_0^2}^{\widehat{s}} dQ^2 \left( \frac{ d\widehat{\sigma}
^{hard}_{ab\rightarrow c(d)}}{dQ^2}\right)\right\}~,
\end{equation}
where the symbols have  their usual meaning. The specific value of 
$p_0$ is fixed by demanding that the energy dependence
of the total cross-section for $pp$ collisions is correctly reproduced
when the expression is convoluted with the structure function of the
nucleons within an eikonal approximation~\cite{G}. 
One way to satisfy this requirement is a scale~\cite{naga}
\begin{equation}
p_0\equiv p_0(\sqrt{s})=0.5(\sqrt{s})^{0.27}
\end{equation}
and we shall use this form, where $\sqrt{s}$ is the nucleon-nucleon
c.m.s.\ energy. 

\medskip

In the two-term Ansatz (1), the hard and soft contributions now have to be 
specified. For the hard collisions above $p_0$ we use the standard form 
\begin{equation}
\frac{d\widehat{\sigma}^{hard}_{ab\rightarrow cd}}{dQ^2}
=\frac{1}{16 \pi \widehat{s}^2}|\overline{M}_{ab\rightarrow cd}
  (\widehat{s},Q^2)|^2
\sim \frac{\pi \alpha_s^2(Q^2)}{Q^2}
\end{equation}
where $|M|^2$ is the process-dependent spin- and colour-averaged squared
matrix element in Born approximation.
To account for higher order contributions, we modify the corresponding cross 
sections by a $K$-factor of 2.5~\cite{bass}. 
 Soft collisions between two partons are assumed to proceed through a 
very low-energy double gluon exchange. This provides a natural continuation 
to the harder collisions above $p_0$, where the dominant one-gluon
exchange processes in $gg\rightarrow gg$, $gq \rightarrow gq$, and
$qq \rightarrow qq$ have the same overall structure~\cite{gustaf}.
A simple and physically plausible form for the soft cross-section
continues the hard cross-section for $Q^2$ below $p_0^2$ down to 
$Q^2=0$ by introducing a regulating term $\beta^2$,
so that we have
\begin{equation}
\frac{d\widehat{\sigma}^{soft}_{ab \rightarrow cd}}{dQ^2}\sim \frac{2\pi
\alpha_s^2(p_0^2)}{Q^2+\beta^2}
\end{equation}
Thus $\beta$  acts as a phenomelogical parameter which governs the overall
magnitude of the integral
 $\sigma^{soft} \sim \ln\left[(p_0^2+\beta^2)/\beta^2\right]$;
it is estimated to be in the range of 0.3 - 1.0 GeV, and we shall use 
$\beta = 0.5$ GeV~\cite{G}.

\medskip

The only further parameter to be specified is the virtuality cut-off 
for the evolution of the time-like partons, below which they are assumed
not to radiate; this is taken to be 1.0 GeV, as determined from 
fits to the hadron production in $e^+e^-$ collisions~\cite{G}.
Finally we add that the maximum possible longitudinal spread for the gluons,
$\Delta z \sim 1/xP$, with $P$ denoting the longitudinal momentum of the 
nucleon, is taken as 1 fm~\cite{bj}. 

\medskip

The parton cascade model provides the phase space distribution of the 
partons. With this given, we have the general form for the colour screening 
mass in the one loop approximation  \cite{BMW,klimov}
\be
\mu^2=-\frac{3\alpha_s}{\pi^2}\,
      \lim_{|\vec{q}|\rightarrow 0}
      \int\, d^3k\,
       \frac{|\vec{k}|}{\vec{q}\cdot\vec{k}}
     \,  \vec{q}\cdot\nabla_{\vec{k}}
   \left[ f_{g}(\vec{k})+\frac{1}{6}\sum_q
    \left\{ f_q(\vec{k})+f_{\overline{q}}(\vec{k})\right\} \right],
\ee
where $\alpha_s$ is the strong coupling constant, the $f_i$ specify
the phase space density of gluons, quarks, and anti-quarks and $q$
runs over the flavour of quarks. It is easy
to verify that in the case of ideal gas of massless partons, where the
$f_i$ reduce to Bose-Einstein or Fermi-Dirac distributions (with
vanishing baryochemical potential $\mu_B$), Eq.\ (5) reduces 
at high temperatures to
\be
\mu^2=4\pi\alpha_s T^2\left(1+\frac{N_f}{6}\right)~.
\ee
On the other hand, for the same system at large $\mu_B$
and low temperature, we get
\begin{equation}
\mu^2=\frac{2}{\pi}\,\alpha_s \mu_B^2 N_f~.
\ee
Following \cite{BMW,sspc}, we shall assume that Eq.\ (5) holds also for the
non-equilibrated partons created by partonic scattering and radiation
in the early stages of the nuclear collision. We
use the parton cascade model to estimate the phase-space density of
the partons. The phase space density of the partons thus
produced can be written as
\begin{equation}
f_{i}(\vec{k})=\frac{2(2\pi)^2}{g_iV}\frac{1}{|\vec{k}|}f_i(k_T,y),
\end{equation}
where $g_i$ gives the relevant degeneracy with $g_g=16$, and
$g_q=g_{\overline{q}}=6$, $V$ is the volume occupied by the partons
at the time $\tau$, and $k_T$ and $y$ are the transverse momenta and
the rapidity of the parton under consideration. The partonic
distribution will be initially anisotropic with respect to the beam axis
and thus the screening mass of a gluon in
such a matter will depend on its direction of propagation.

\medskip

It is known that the initial distribution of the partons in the
parton cascade model resembles a plateau extending to rapidities
$\pm Y$. $Y$ is found to be of the order of 2 at $\sqrt{s}=20$ A$\cdot$GeV,
about 3 at $\sqrt{s}=$ 200 A$\cdot$GeV, and more than 4--5 at energies
likely to be attained at LHC. To allow a comparison of the
evolving scenarios as the energy is increased, we shall keep $Y=2$
in the following. The step-wise development of the parton-cascade
in time as implemented in Monte-Carlo simulation VNI~\cite{vni}
is of great help in accounting for partons at some given
instant of time, in a given volume. It may be noted that in the
parton cascade model, the partons are given a formation-time
and are counted as real particles only after this
time. For ease of computation, we approximate the parton distributions
as
\begin{equation}
f_i(k_T,y)=\frac{1}{2Y}f_i(k_T)\left[\theta(y+Y)-\theta(y-Y)\right],
\end{equation}
where $f_i(k_T)=dN_i/dk_T^2$ for the partons in the volume we choose
for estimating the screening mass. We are interested in seeing the
variation with time $\tau$ and with distance $r_T$ in a plane transverse 
to the collision axis; in central collisions, the density of partons 
is highest along this axis.  
We shall therefore choose different zones $R_i<r_T<R_{i+1}$ 
in the central slice near $z = 0$. The screening masses in central $AA$
collisions are then given by \cite{BMW,sspc}
\begin{eqnarray}
\mu_T^2&=&\frac{48 \alpha_s}{\tau (R_{i+1}^2-R_i^2)}\left(
\frac{\sin^{-1}(\tanh Y)}{Y}\right)
\int dk_T \left[\frac{1}{g_G}f_g(k_T)+\frac{1}{g_Q}\sum_f \left\{
f_q(k_T)+
f_{\overline{q}}(k_T)\right\}\right] \nonumber\\
& &\\
\mu_{\parallel}^2&=&\mu_T^2\left(1+\frac{1}{\sinh Y \sin^{-1}(\tanh Y)}\right).
\end{eqnarray}
for the transverse and longitudinal directions, respectively. For $Y=2$, 
the factor in brackets in Eq.\ (11) is about 1.2, so that $\mu_T \simeq 
\mu_{||}$ within 10 \%, and hence we shall in the remainder 
consider only $\mu_T\equiv \mu$.

\medskip

The evolution time slice of interest is somewhat arbitrary; it can probably 
be fixed more precisely when discussing specific signatures, and we shall 
come back to this point a little further on. The calculations to be shown 
here will start at proper time $\tau = 0.1$ fm, assuming that at least this 
much time is needed to establish any kind of medium. For $\tau \geq 0.5$ fm, 
the screening masses tend to become only weakly time-dependent, so we 
present calculations up to about 1 fm. 

\medskip

We begin by calculating the screening mass for partonic matter in central 
$S - S$ collisions at SPS, RHIC, and LHC energies,
for transverse distances $r_T\leq$ 2 fm (Fig.~1).
Here and in all subsequent calculations, the cascade starts when the 
two incoming nuclei are centered at $z=\pm 1$ fm and $\tau=-1$ fm.
At SPS energy, the screening mass is found to be always less than 0.25 GeV and 
thus considerably below the deconfinement point $\mu_c \simeq 0.5$ GeV. We 
therefore expect that the partonic matter produced in this case could not have
been deconfined. In contrast, RHIC energies bring $\mu$ just into the 
deconfinement zone, and for LHC, $\mu > > \mu_c$ for all times considered here,
so that at the LHC, all collisions at reasonable impact parameters will 
lead deeply into the deconfinement regime. 

\medskip

Next we turn to $Pb - Pb$ collisions at the SPS, see Fig.~2. 
Here there is a hot center, i.e., at early times, a region for sufficiently
small $r_T$ reaches screening masses around $\mu_c$. For $Au - Au$ 
collisions at RHIC (Fig.~3), $\mu$ increases by more than a factor three, 
so that now essentially the entire collision volume falls into the 
deconfined region. To study the onset of deconfinement at RHIC would thus 
seem to require smaller nuclei, and so we show in Fig.~4 corresponding 
results for $Cu - Cu$ collisions, confirming this expectation. 

\medskip

Before considering applications of these results to the study of 
deconfinement signals, we should note that our calculations are based on 
perturbative partonic interactions and a phenomenological extension thereof 
into the soft regime. There are indications \cite{kajantie}
that non-perturbative corrections may lead to considerably larger screening 
masses, and it would clearly be of great interest to obtain more precise
results from finite temperature lattice QCD.  

\medskip

For a test of deconfinement in nuclear collisions, the production and 
suppression of quarkonium states appears so far to be the most suitable 
probe \cite{MS}.
The tight binding of most of these states prevents their dissociation in 
confined matter \cite{KS3};
in a deconfined medium, a hierarchy of suppression is predicted, 
governed by the size or the binding energy of the state in 
question \cite{KAR}. 
In Table 1, we summarize the screening masses obtained in potential theory
based on a screened confining potential \cite{KAR}. 

\medskip

\begin{table}[h]
\centerline{
\begin{tabular}{|l|l|l|l|l|l|} 
\hline
  & & & & &   \\
 State & $J/\psi$ & $\chi_c$ &$\Upsilon$ &$\Upsilon^\prime$ & $\chi_b$ \\
  & & & & &   \\
\hline
& & & & & \\ 
$\mu^{\rm diss}_x$ (GeV) & 0.70 & 0.34 & 1.57 & 0.67 & 0.56 \\
& & & & &  \\
\hline
\end{tabular}
}
\end{table}
\centerline{Table 1: Critical screening masses for quarkonia.}

\bigskip

From Fig.~2 we see that at the SPS, collisions of sufficiently heavy 
nuclei reach a regime in which much of the $\chi_c$ production should be 
suppressed by deconfinement. The effect on direct $J / \psi$ production seems 
to be a more quantitative question - slight non-perturbative contributions
could perhaps enhance $\mu$ enough to reach deconfinement values. In 
$Au - Au$ collisions at RHIC, essentially all charmonium production should be 
suppressed, see Fig.~3, and if sufficient statistics become possible, the 
onset of bottonium suppression could be studied here for the first time. The 
LHC, finally, would seem to be an ideal tool to carry out a 
systematic study of bottonium suppression. 

\medskip

We thus find that the effect of screening mass considerations in the parton 
cascade model leads to a pattern of quarkonium suppression which is very 
similar to what has been obtained in other, more global approaches \cite{Hwa}.
In particular, recent suppression studies based on string \cite{Armesto}
or parton percolation \cite{Satz-P,Satz-T} 
agree very nicely with the present results. An open question which could be 
addressed in the present, time-dependent approach is the dissociation of a 
charmonium state in an evolving medium. As a first guess, we have assumed that 
such states are dissociated once the screening length reaches 
deconfinement values. For more quantitative estimates, one should solve the 
bound-state problem in the presence of time-dependent screening, to see how 
long the nascent charmonium state has to spend in a deconfining medium before
it is fully dissociated.

\medskip

In closing we want to emphasize that our main aim was to outline how 
a microscopic evolution scheme based on parton interactions can be used
to study the onset of deconfinement and its effect on deconfinement signatures.
We have here used the parton cascade model; it would clearly be of interest 
to see if other, similar or not so similar models corroborate our results.   

\bigskip

\centerline{\bf Acknowledgments}

\medskip

One of us (DKS) gratefully acknowledges the hospitality of University of
Bielefeld where part of this work was done. He would also like to acknowledge
useful discussions with Avijit Ganguly, Munshi Golam Mustafa, 
Bikash Sinha, and Markus Thoma.

\newpage
\newpage
\newpage
\begin{figure}
\psfig{file=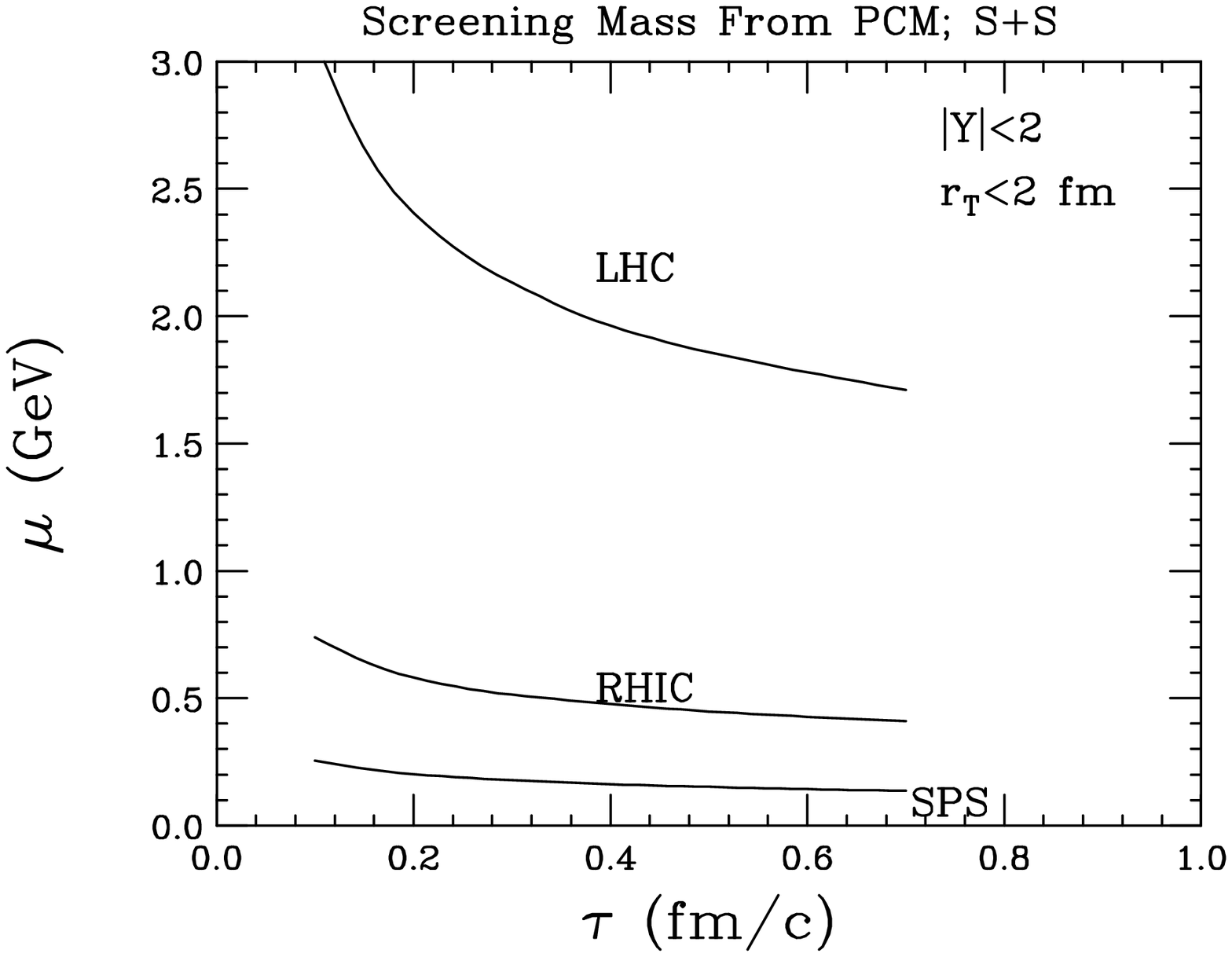,height=12cm,width=15cm}
\vskip 0.1in
\caption{The time evolution of the colour
screening mass in central $S-S$ collisions 
at SPS, RHIC and LHC energies.}
\end{figure}
\newpage

\begin{figure}
\psfig{file=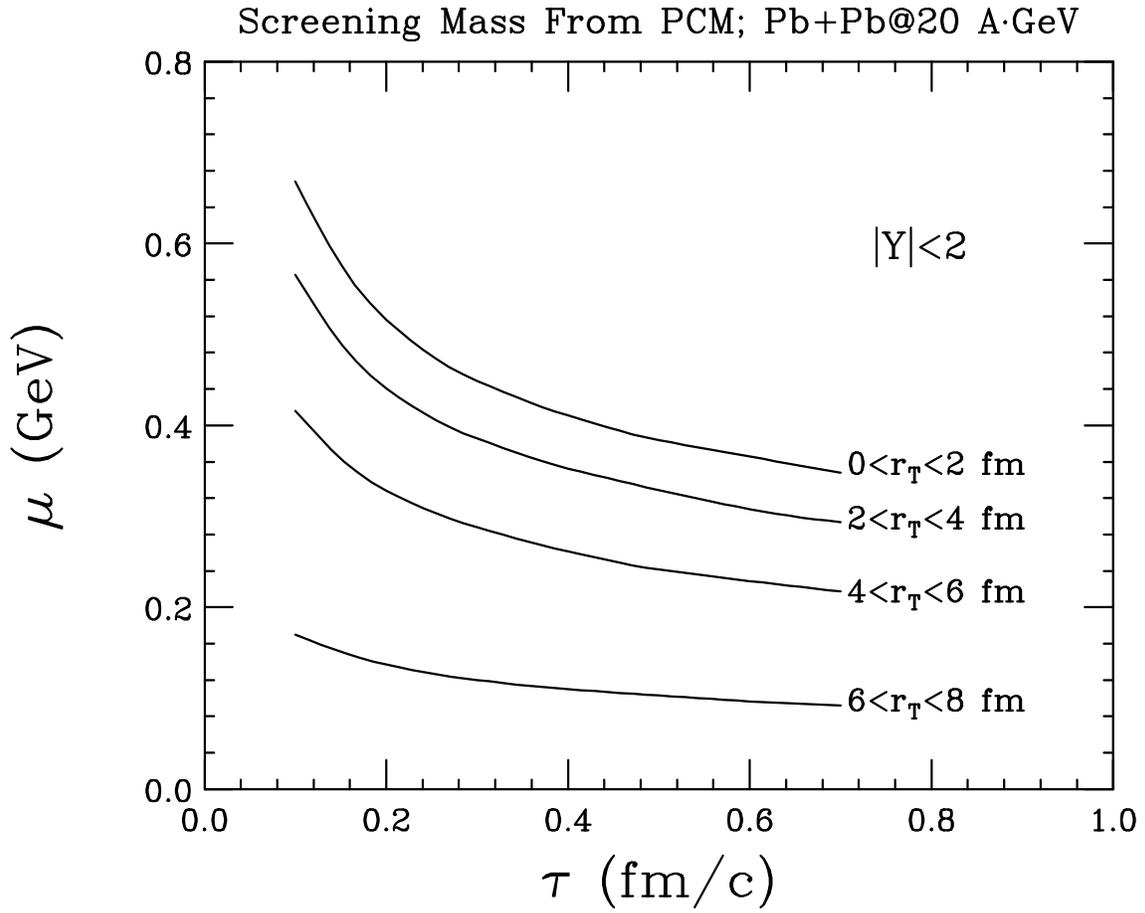,height=12cm,width=15cm}
\vskip 0.1in
\caption{The time evolution of the colour
screening mass in central $Pb-Pb$ collisions at SPS energy
in different radial zones.}
\end{figure}
\newpage

\begin{figure}
\psfig{file=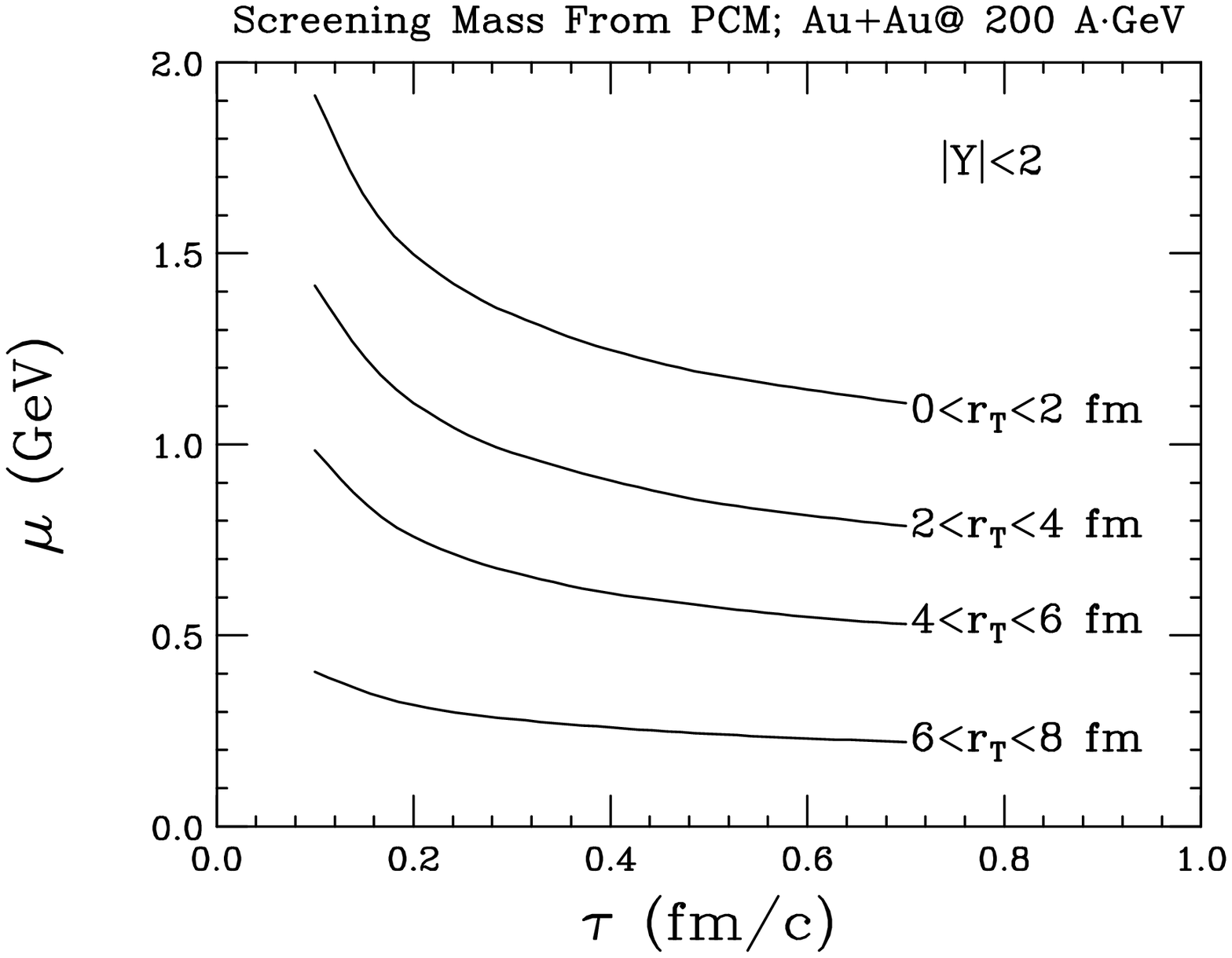,height=12cm,width=15cm}
\vskip 0.1in
\caption{The time evolution of the colour
screening mass in central $Au-Au$ collisions at RHIC energy
in different radial zones.}
\end{figure}
\newpage
\begin{figure}
\psfig{file=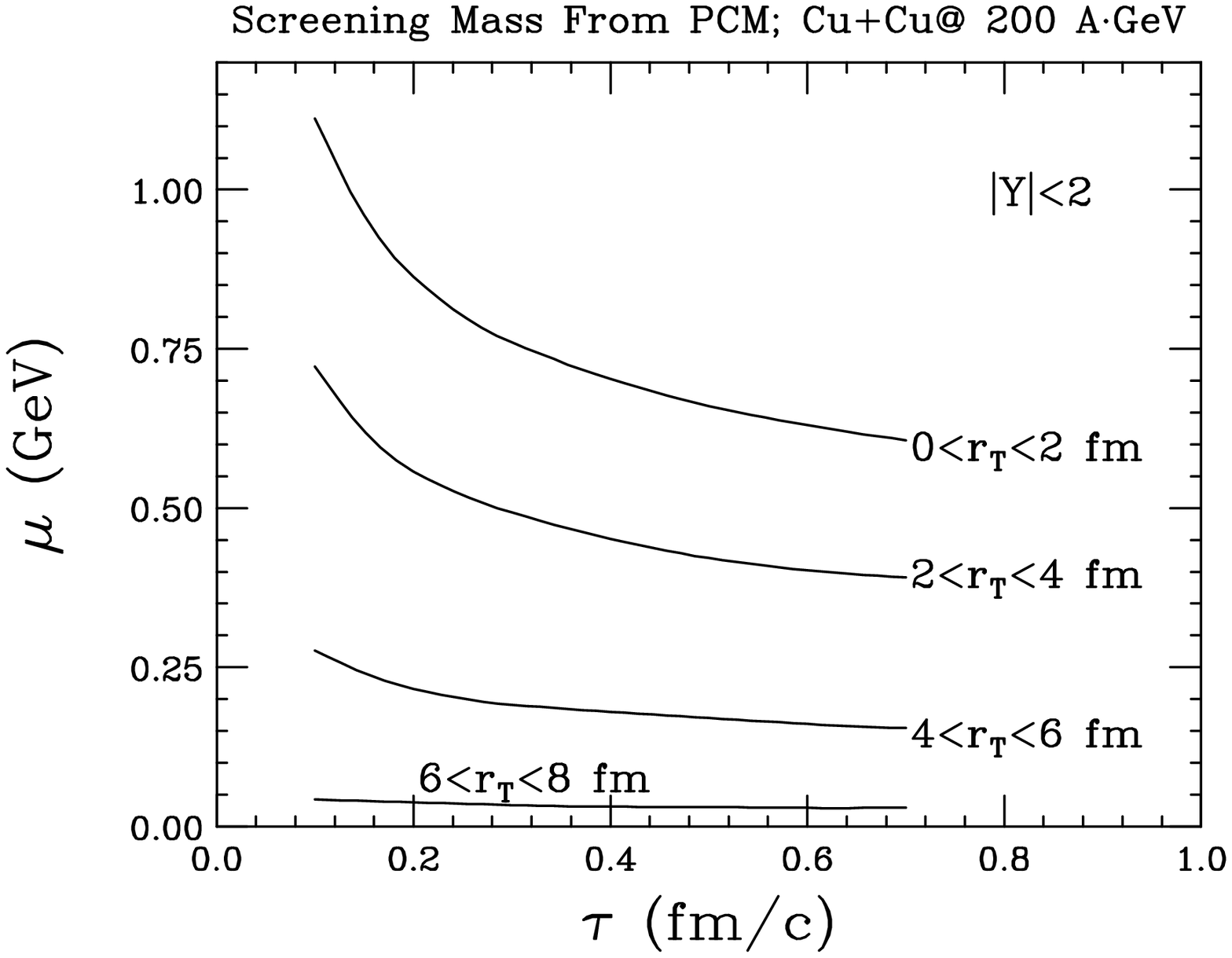,height=12cm,width=15cm}
\vskip 0.1in
\caption{The time evolution of the colour
screening mass in central $Cu-Cu$ collisions at RHIC energy
in different radial zones.}
\end{figure}

\bigskip

\begin{thebibliography}{37}

\bibitem{GM} K.\ Geiger and B.\ M\"uller, Nucl.\ Phys.\ B 369 (1992) 600.


\bibitem{G}
K.\ Geiger, Phys.\ Rep.\ 258 (1995) 376 and references there-in.

\bibitem{BMW} T.\ S.\ Bir\'o, B.\ M\"{u}ller, and X.-N.\ Wang, \PRL B
283 (1992) 171.

\bibitem{sspc} K.\ J.\ Eskola, B.\ M\"{u}ller, and X.-N. Wang, Phys. Lett.
B  374 (1996) 20.

\bibitem{Armesto} N.\ Armesto et al., \PRL 77 (1996) 3736.


\bibitem{Satz-P} M.\ Nardi and H.\ Satz, Phys. Lett. B 442 (1998) 14.

\bibitem{Satz-T} H.\ Satz, hep-ph/9908339.

\bibitem{lattice} U.\ M.\ Heller, F.\ Karsch, and J.\ Rank,
Phys. Lett. B 355 (1995) 511.

\bibitem{eg} J.\ Ellis and K.\ Geiger, \PR D 52 (1995) 1500;
J.\ Ellis and K.\ Geiger, \PR D 54 (1996) 1967.

\bibitem{naga} N.\ Abou-El-Naga, K.\ Geiger, and B.\ M\"{u}ller,
               J.\ Phys.\ G 18 (1992) 797.


\bibitem{bass} S.\ A.\ Bass and B.\ M\"{u}ller, nucl-th/9908014.

\bibitem{gustaf} G.\ Gustafson, Z.\ Phys.\ C 15 (1982) 155.

\bibitem{bj}   J.\ D.\ Bjorken in ``Current Induced reactions'', Proc.
Int. Summer Institute on Theoretical Particle Physics, Hamburg 1975,
J. K\"{o}rner et al.\ (Eds.), {\sl Lecture Notes in Physics}, Vol. 56, 
Springer, New York, 1976.

\bibitem{klimov} O.\ K.\ Kalashnikov and V.\ V.\ Klimov, Sov.\ J.\ 
Nucl.\ Phys.\ 31 (1980) 699; V.\ V.\ Klimov, Sov. Phys. JETP  55 (1982) 199.

\bibitem{vni} K.\ Geiger, Comp. Phys. Comm. 104 (1997) 70;
K.\ Geiger, R.\ Longacre, and D.\ K.\ Srivastava, nucl-th/9806102.
The results reported here have been obtained using the revised
version VNI/BMS by S.\ A.\ Bass, B.\ M\"{u}ller, and D.\ K.\ Srivastava 
(to be published ), which removes several 
inconsistencies in earlier versions of the code.

\bibitem{kajantie} K.\ Kajantie et al., \PRL 17 (1997) 3130.

\bibitem{MS} T.\ Matsui and H.\ Satz, \PL B 178 (1986) 416.

\bibitem{KS3} D.\ Kharzeev and H.\ Satz, \PL B 334 (1994) 155.

\bibitem{KAR} F.\ Karsch and H.\ Satz, Z. Phys. C 51 (1991) 209.


\bibitem{Hwa} See e.g., D.\ Kharzeev and H.\ Satz in R.\ Hwa (Ed.),
{\sl Quark-Gluon Plasma 2}, World Scientific, Singapore 1995. 

\end{thebibliography}
\end{document}